\documentclass[10pt]{article}

\newlength{\minitwocolumn}
\font\teneufm=eufm10
\font\seveneufm=eufm7
\font\fiveeufm=eufm5
\newfam\eufmfam
\textfont\eufmfam=\teneufm
\scriptfont\eufmfam=\seveneufm
\scriptscriptfont\eufmfam=\fiveeufm

\makeatletter
\@addtoreset{equation}{section}
\makeatother

\title{\large{\bf BOSONIZATION AND VERTEX OPERATOR OF\\ 
SUPERSYMMETRY $U_q(\widehat{sl}(N|1))$ 
FOR LEVEL $k$}}

\begin{document}

\maketitle

\begin{center}
{TAKEO KOJIMA}
\\~\\

{\it
Faculty of Engineering,
Yamagata University, Jonan 4-3-16, Yonezawa 992-8510,
Japan\\
kojima@yz.yamagata-u.ac.jp}
\end{center}

~\\~\\

\begin{abstract}
We construct a bosonization of the quantum superalgebra $U_q(\hat{sl}(N|1))$
for an arbitrary level $k$.  
We construct the screening 
that commutes with the quantum superalgebra
for an arbitrary level $k \neq -N+1$.
We propose a bosonization of the vertex operator that gives the intertwiner 
among the Wakimoto realization and the typical representation. 
\end{abstract}

~\\~\\

\section{Introduction}
There have been many developments in exactly solvable models.
Various methods were invented to solve models.
The bosonization provides a powerful method to study 
exactly solvable models.
We review recent developments in bosonizations
of the $U_q(\widehat{sl}(N|1))$ \cite{Kojima1, Kojima2, Kojima3}.
The trace of our bosonizations of the vertex operators
gives the correlation function of the higher level(spin) $k$ and rank $N$
generalization of the supersymmetric $t$-$J$ model ($U_q(\widehat{sl}(2|1))$
for level $k=1$) \cite{ZR}.

\section{Bosonization of $U_q(\widehat{sl}(N|1))$}

We recall Drinfeld generators of
the superalgebra
$U_q(\widehat{sl}(N|1))$ \cite{Yamane}.
We fix a complex number $q \neq 0, |q|<1$.
We use the notations $[x,y]=xy-yx,~\{x,y\}=xy+yx,~[a]_q=\frac{q^a-q^{-a}}{
q-q^{-1}}$.
The Cartan matrix 
$(A_{i,j})_{0\leq i,j \leq N}$
of the affine Lie algebra $\widehat{sl}(N|1)$ is
given by
\begin{eqnarray}
A_{i,j}=
(\nu_i+\nu_{i+1})\delta_{i,j}-
\nu_i \delta_{i,j+1}-\nu_{i+1}\delta_{i+1,j}.
\nonumber
\end{eqnarray}
Here we set $\nu_1=\cdots =\nu_N=+, \nu_{N+1}=\nu_0=-$.
The Drinfeld generators of
the quantum
superalgebra $U_q(\widehat{sl}(N|1))$
are
$x_{i,m}^\pm,~h_{i,m},~c,~(1\leq i \leq N,
m \in {\bf Z})$.
Defining relations are
\begin{eqnarray}
&&~c : {\rm central},~[h_i,h_{j,m}]=0,\nonumber
\\
&&~[a_{i,m},h_{j,n}]=\frac{[A_{i,j}m]_q[cm]_q}{m}q^{-c|m|}
\delta_{m+n,0}~~(m,n\neq 0),~[h_i,x_j^\pm(z)]=\pm A_{i,j}x_j^\pm(z),
\nonumber
\\
&&~[h_{i,m}, x_j^+(z)]=\frac{[A_{i,j}m]_q}{m}
q^{-c|m|} z^m x_j^+(z),~~[h_{i,m}, x_j^-(z)]=-\frac{[A_{i,j}m]_q}{m}
z^m x_j^-(z)~~(m \neq 0),
\nonumber
\\
&&(z_1-q^{\pm A_{i,j}}z_2)
x_i^\pm(z_1)x_j^\pm(z_2)
=
(q^{\pm A_{j,i}}z_1-z_2)
x_j^\pm(z_2)x_i^\pm(z_1)~~~{\rm for}~|A_{i,j}|\neq 0,
\nonumber
\\
&&
x_i^\pm(z_1)x_j^\pm(z_2)
=
x_j^\pm(z_2)x_i^\pm(z_1)~~~{\rm for}~|A_{i,j}|=0, (i,j)\neq (N,N),
~~\{x_N^\pm(z_1), x_N^\pm(z_2)\}=0,
\nonumber\\
&&~[x_i^+(z_1),x_j^-(z_2)]
=\frac{\delta_{i,j}}{(q-q^{-1})z_1z_2}
\left(
\delta(q^{-c}z_1/z_2)\psi_i^+(q^{\frac{c}{2}}z_2)-
\delta(q^{c}z_1/z_2)\psi_i^-(q^{-\frac{c}{2}}z_2)
\right), \nonumber\\
&& ~~~~~{\rm for}~~(i,j) \neq (N,N),\nonumber\\
&&~\{x_N^+(z_1),x_N^-(z_2)\}
=\frac{1}{(q-q^{-1})z_1z_2}
\left(
\delta(q^{-c}z_1/z_2)\psi_N^+(q^{\frac{c}{2}}z_2)-
\delta(q^{c}z_1/z_2)\psi_N^-(q^{-\frac{c}{2}}z_2)
\right), \nonumber\\
&& 
\left(
x_i^\pm(z_{1})
x_i^\pm(z_{2})
x_j^\pm(z)-(q+q^{-1})
x_i^\pm(z_{1})
x_j^\pm(z)
x_i^\pm(z_{2})
+x_j^\pm(z)
x_i^\pm(z_{1})
x_i^\pm(z_{2})\right)\nonumber\\
&&+\left(z_1 \leftrightarrow z_2\right)=0
~~~{\rm for}~|A_{i,j}|=1,~i\neq N,\nonumber
\end{eqnarray}
where we have used
$\delta(z)=\sum_{m \in {\bf Z}}z^m$.
Here we have
used the generating function
$x_j^\pm(z)=
\sum_{m \in {\bf Z}}x_{j,m}^\pm z^{-m-1}$,
$\psi_i^\pm(q^{\pm \frac{c}{2}}z)=q^{\pm h_i}
e^{
\pm (q-q^{-1})\sum_{m>0}h_{i,\pm m}z^{\mp m}}$
and the abbreviation $h_i=h_{i,0}$.

We construct bosonizations of superalgebra $U_q(\hat{sl}(N|1))$
for an arbitrary level \cite{Kojima1}. 
We fix the level $c=k \in {\bf C}$.
We introduce the bosons
and the zero-mode operators
$a_m^j, Q_a^j$ $(m \in {\bf Z},
1\leq j \leq N)$, 
$b_m^{i,j}, Q_b^{i,j}$
$(m \in {\bf Z}, 1\leq i<j \leq N+1)$,
$c_m^{i,j}, Q_c^{i,j}$
$(m \in {\bf Z}, 1\leq i<j \leq N)$
which satisfy
\begin{eqnarray}
&&~[a_m^i,a_n^j]=\frac{[(k+N-1)m]_q[A_{i,j}m]_q}{m}
\delta_{m+n,0}~~~(m,n \neq 0),~~~[a_0^i, Q_a^j]=(k+N-1)A_{i,j},\nonumber
\\
&&~[b_m^{i,j},b_n^{i',j'}]=
-\nu_i \nu_j \frac{[m]_q^2}{m}
\delta_{i,i'}\delta_{j,j'}\delta_{m+n,0}~~~(m,n \neq 0),~~~
[b_0^{i,j},Q_b^{i',j'}]=
-\nu_i \nu_j \delta_{i,i'}\delta_{j,j'},\nonumber
\\
&&~[c_m^{i,j},c_n^{i',j'}]=
\frac{[m]_q^2}{m}
\delta_{i,i'}\delta_{j,j'}
\delta_{m+n,0}~~~(m,n \neq 0),~~~
[c_0^{i,j},Q_c^{i',j'}]=
\delta_{i,i'}\delta_{j,j'},\nonumber
\\
&&~[Q_b^{i,j},Q_b^{i',j'}]=\delta_{j,N+1}\delta_{j',N+1}
\pi \sqrt{-1}~~~~~((i,j) \neq (i',j')).\nonumber
\end{eqnarray}
Other commutation relations are zero.
In what follows we use the standard symbol of the normal orderings $: :$
and use the following abbreviations 
$b^{i,j}(z), c^{i,j}(z), b_\pm^{i,j}(z),
a^j_\pm(z)$ and $\left(\frac{\gamma_1}{\beta_1}\frac{\gamma_2}{\beta_2}
\cdots \frac{\gamma_r}{\beta_r}
~a^i \right)\left(z|\alpha \right)$ given by
\begin{eqnarray}
b^{i,j}(z)=
-\sum_{m \neq 0}\frac{b_m^{i,j}}{[m]_q}z^{-m}+Q_b^{i,j}+b_0^{i,j}{\rm log}z,~
c^{i,j}(z)=
-\sum_{m \neq 0}\frac{c_m^{i,j}}{[m]_q}z^{-m}+Q_c^{i,j}+c_0^{i,j}{\rm log}z,
\nonumber
\\
b_\pm^{i,j}(z)=\pm (q-q^{-1})\sum_{\pm m>0}b_m^{i,j} 
z^{-m} \pm b_0^{i,j}{\rm log}q,~
a_\pm^{j}(z)=\pm (q-q^{-1})\sum_{\pm m>0}a_m^{j} 
z^{-m}\pm a_0^j {\rm log}q,\nonumber\\
\left(\frac{\gamma_1}{\beta_1}\frac{\gamma_2}{\beta_2}
\cdots \frac{\gamma_r}{\beta_r}
~a^i \right)\left(z|\alpha \right)=
-\sum_{m \neq 0}\frac{[\gamma_1 m]_q \cdots [\gamma_r m]_q}
{[\beta_1 m]_q \cdots [\beta_r m]_q}
\frac{a^i_m}{[m]_q} q^{-\alpha |m|}z^{-m}
+\frac{\gamma_1 \cdots \gamma_r}{\beta_1 \cdots \beta_r}
(Q_a^i+a_0^i {\rm log}z).\nonumber
\end{eqnarray}
The generating functions 
$x_i^\pm(z)$,
$\psi_i^\pm(z)$, $(1\leq i \leq N)$
of $U_q(\widehat{sl}(N|1))$ 
for an arbitrary level $k$ are realized by
the bosonic operators as follows. This is main result of \cite{Kojima1}.
\begin{eqnarray}
x_i^+(z)&=&
\frac{1}{(q-q^{-1})z}
:\sum_{j=1}^i
e^{(b+c)^{j,i}(q^{j-1}z)+\sum_{l=1}^{j-1}
(b_+^{l,i+1}(q^{l-1}z)-b_+^{l,i}(q^lz))}
\times \nonumber
\\
&\times&
\left\{
e^{b_+^{j,i+1}(q^{j-1}z)-
(b+c)^{j,i+1}(q^jz)}-
e^{b_-^{j,i+1}(q^{j-1}z)-
(b+c)^{j,i+1}(q^{j-2}z)}\right\}:,\nonumber
\\
x_N^+(z)
&=&:
\sum_{j=1}^N 
e^{(b+c)^{j,N}(q^{j-1}z)
+b^{j,N+1}(q^{j-1}z)
-\sum_{l=1}^{j-1}(b_+^{l,N+1}(q^lz)+b_+^{l,N}(q^lz))}:,
\nonumber
\\
x_i^-(z)&=&
q^{k+N-1}
:e^{a_+^i(q^{\frac{k+N-1}{2}}z)
-b^{i,N+1}(q^{k+N-1}z)-b_+^{i+1,N+1}(q^{k+N-1}z)+b^{i+1,N+1}(q^{k+N}z)}:
\nonumber
\\
&+&
\frac{1}{(q-q^{-1})z}:
\left\{
\sum_{j=1}^{i-1}
e^{
a_-^i(q^{-\frac{k+N-1}{2}}z)
+(b+c)^{j,i+1}(q^{-k-j}z)
+b_-^{i,n+1}(q^{-k-n}z)-b_-^{i+1,n+1}(q^{-k-n+1}z)}
\right.
\nonumber\\
&\times&
e^{\sum_{l=j+1}^i 
(b_-^{l,i+1}(q^{-k-l+1}z)-b_-^{l,i}(q^{-k-l}z))
+\sum_{l=i+1}^N
(b_-^{i,l}(q^{-k-l}z)-b_-^{i+1,l}(q^{-k-l+1}z))}\nonumber\\
&\times&
\left(
e^{-b_-^{j,i}(q^{-k-j}z)
-(b+c)^{j,i}(q^{-k-j+1}z)}
-
e^{-b_+^{j,i}(q^{-k-j}z)
-(b+c)^{j,i}(q^{-k-j-1}z)}
\right)
\nonumber\\
&+&
e^{a_-^i(q^{-\frac{k+N-1}{2}}z)+(b+c)^{i,i+1}(q^{-k-i}z)
+\sum_{l=i+1}^N(b_-^{i,l}(q^{-k-l}z)
-b_-^{i+1,l}(q^{-k-l+1}z))
+b_-^{i,N+1}(q^{-k-N}z)-b_-^{i+1,N+1}(q^{-k-N+1}z)}
\nonumber
\\
&-&
e^{a_+^i(q^{\frac{k+N-1}{2}}z)
+(b+c)^{i,i+1}(q^{k+i}z)
+\sum_{l=i+1}^N(b_+^{i,l}(q^{k+l}z)
-b_+^{i+1,l}(q^{k+l-1}z))
+b_+^{i,N+1}(q^{k+N}z)-b_+^{i+1,N+1}(q^{k+N-1}z)}
\nonumber\\
&-&
\sum_{j=i+1}^{N-1}
e^{a_+^i(q^{\frac{k+N-1}{2}}z)+
(b+c)^{i,j+1}(q^{k+j}z)
+b_+^{i,N+1}(q^{k+N}z)-b_+^{i+1,N+1}(q^{k+N-1}z)
+\sum_{l=j+1}^N
(b_+^{i,l}(q^{k+l}z)-b_+^{i+1,l}(q^{k+l-1}z))}\nonumber\\
&\times&
\left.
\left(
e^{b_+^{i+1,j+1}(q^{k+j}z)-(b+c)^{i+1,j+1}(q^{k+j+1}z)}
-
e^{b_-^{i+1,j+1}(q^{k+j}z)-(b+c)^{i+1,j+1}(q^{k+j-1}z)}
\right)\right\}:.
\nonumber\\
x_N^-(z)&=&
\frac{1}{(q-q^{-1})z}
:\left\{
\sum_{j=1}^{N-1}
e^{
a_-^N(q^{-\frac{k+N-1}{2}}z)
-b_+^{j,N+1}(q^{-k-j}z)-b^{j,N+1}(q^{-k-j-1}z)
-\sum_{l=j+1}^{N-1}(b_-^{l,N}(q^{-k-l}z)
+b_-^{l,N+1}(q^{-k-l}z))}
\right.
\nonumber\\
&\times&
\left.
q^{j-1}\left(
e^{-b_+^{j,N}(q^{-k-j}z)-(b+c)^{j,N}(q^{-k-j-1}z)}
-
e^{
-b_-^{j,N}(q^{-k-j}z)-(b+c)^{j,N}(q^{-k-j+1}z)}
\right)\right\}:
\nonumber\\
&+&
q^{N-1}:\left(
e^{
a_+^N(q^{\frac{k+N-1}{2}}z)-b^{N,N+1}(q^{k+N-1}z)}
-
e^{
a_-^N(q^{-\frac{k+N-1}{2}}z)-b^{N,N+1}(q^{-k-N+1}z)}
\right):.
\nonumber
\end{eqnarray}
\begin{eqnarray}
\psi_i^\pm(q^{\pm \frac{k}{2}}z)&=&
e^{a_\pm^i(q^{\pm \frac{k+N-1}{2}}z)+
\sum_{l=1}^i(b_\pm^{l,i+1}(q^{\pm(l+k-1)}z)-b_\pm^{l,i}
(q^{\pm(l+k)}z)}\nonumber
\\
&&\times
e^{\sum_{l=i+1}^{N}(b_\pm^{i,l}(q^{\pm(k+l)}z)-
b_\pm^{i-1,l}(q^{\pm(k+l-1)}z)
+b_\pm^{i,N+1}(q^{\pm(k+N)}z)-
b_\pm^{i+1,N+1}(q^{\pm(k+N-1)}z)},
\nonumber
\\
\psi_N^\pm(q^{\pm \frac{k}{2}}z)
&=&
e^{a_\pm^N(q^{\pm \frac{k+N-1}{2}}z)-
\sum_{l=1}^{N-1}
(b_\pm^{l,N}(q^{\pm (k+l)}z)
+b_\pm^{l,N+1}(q^{\pm (k+l)}z))}.\nonumber
\end{eqnarray}

\section{Vertex Operator}

We construct the screening ${\cal Q}_i$ 
that commutes with the quantum superalgebra
$U_q(\widehat{sl}(N|1))$
for an arbitrary level $k \neq -N+1$ \cite{Kojima3}.
The Jackson integral with parameter $p \in {\bf C}$ $(|p|<1)$
and $s \in {\bf C}^*$ is defined by
$\int_0^{s \infty}f(z)d_pz=s(1-p)\sum_{m \in {\bf Z}}f(sp^m)p^m$.
We introduce the screening operators 
${\cal Q}_i$ $(1\leq i \leq N)$ 
by using the Jackson integral.
This is one of main result of \cite{Kojima3}.
\begin{eqnarray}
{\cal Q}_i=\int_0^{s \infty}
:e^{-\left(\frac{1}{k+N-1}a^i\right)
\left(z\left|\frac{k+N-1}{2}\right.\right)}
\widetilde{S}_i(z):d_pz,~~~(p=q^{2(k+N-1)}).\nonumber
\end{eqnarray}
Here we have set
the bosonic operators 
$\widetilde{S}_i(z)$ $(1\leq i \leq N)$ by
\begin{eqnarray}
\widetilde{S}_i(z)&=&\frac{1}{(q-q^{-1})z}\sum_{j=i+1}^N
:\left(
e^{-b_-^{i,j}(q^{N-1-j}z)-(b+c)^{i,j}(q^{N-j}z)}
-
e^{-b_+^{i,j}(q^{N-1-j}z)-(b+c)^{i,j}(q^{N-j-2}z)}\right)\nonumber\\
&^\times&
e^{(b+c)^{i+1,j}(q^{N-1-j}z)+\sum_{l=j+1}^N
(b_-^{i+1,l}(q^{N-l}z)-b_-^{i,l}(q^{N-l-1}z))
+b_-^{i+1,N+1}(z)-b_-^{i,N+1}(q^{-1}z)}:
\nonumber\\
&+&q:e^{b^{i,N+1}(z)
+b_+^{i+1,N+1}(z)-b^{i+1,N+1}(qz)}:
~~~~~(1\leq i \leq N-1),
\nonumber\\
\widetilde{S}_N(z)&=&-q^{-1}:e^{b^{N,N+1}(z)}:.\nonumber
\end{eqnarray}
The screening ${\cal Q}_i$ commutes with the quantum superalgebra.
\begin{eqnarray}
[{\cal Q}_i, U_q(\widehat{sl}(N|1))]=0~~~(1\leq i \leq N).\nonumber
\end{eqnarray}

For $p_a^i \in {\bf C}$ $(1\leq i \leq N)$,
$p_b^{i,j} \in {\bf C}$ $(1\leq i<j \leq N+1)$,
$p_c^{i,j} \in {\bf C}$ $(1\leq i<j \leq N)$,
we set
the vector $|p_a, p_b, p_c \rangle$ which satisfies
the following conditions.
\begin{eqnarray}
&&a_m^i|p_a,p_b,p_c\rangle=
b_m^{i,j}|p_a,p_b,p_c\rangle=
c_m^{i,j}|p_a,p_b,p_c\rangle=0~~~(m>0),\nonumber\\
&&a_0^i|p_a,p_b,p_c\rangle=p_a^i |p_a,p_b,p_c\rangle,~
b_0^{i,j}|p_a,p_b,p_c\rangle=p_b^{i,j} |p_a,p_b,p_c\rangle,~
c_0^{i,j}|p_a,p_b,p_c\rangle=p_c^{i,j} |p_a,p_b,p_c\rangle.
\nonumber
\end{eqnarray}
The boson Fock space $F(p_a,p_b,p_c)$
is generated by
the bosons $a_m^i, b_m^{i,j}, c_m^{i,j}$
on the vector $|p_a,p_b,p_c\rangle$.
We set the space $F(p_a)$ by
\begin{eqnarray}
F(p_a)=
\bigoplus
_{
p_b^{i,j}=-p_c^{i,j} \in {\bf Z}~(1\leq i<j \leq N)
\atop{
p_b^{i,N+1} \in {\bf Z}~(1\leq i \leq N)
}}F(p_a,p_b,p_c).\nonumber
\end{eqnarray}
We would like to construct the vertex operator 
${\Phi}^*(z)$ \cite{Kojima3}
which gives the intertwiner 
among $F(p_a)$ and the typical representation \cite{PT}.
\begin{eqnarray}
{\Phi}^*(z): {F}(p_a) \longrightarrow 
{F}(p_a+l_a+x_a) \otimes V_z^{* S},~~~
{\Phi}^*(z)=\sum_{j}\Phi_j^*(z)\otimes v_j^*.\nonumber 
\end{eqnarray}
Here $V_z^{* S}$
is the dual evaluation representation of the typical representation
with the weight 
$l_a=(l_a^1,l_a^2,\cdots,l_a^N)$ \cite{PT}.
The basis of $V^{* S}$ is given by $\{v_j^*\}$.
The coefficients are linear maps :
${\Phi}_j^*(z): {F}(p_a+l_a+x_a) \longrightarrow {F}(p_a)$.
For $l_a=(l_a^1,l_a^2,\cdots,l_a^N)$ and $x_a=(x_a^1,x_a^2,\cdots,x_a^N)$,
we set the highest element $\Phi_1^*(z)$ of the vertex operator by
\begin{eqnarray}
\Phi_1^*(z)=
:\prod_{j=1}^N {\cal Q}_j^{x_j}~~
e^{-\sum_{i,j=1}^N
\left(\frac{l_a^i}{k+N-1}
\frac{{\rm Min}(i,j)}{N-1}\frac{N-1-{\rm Max}(i,j)}{1}
a^j\right)(q^kz|-\frac{k-N+1}{2})}
:,\nonumber
\end{eqnarray}
where $(x_1,x_2,\cdots,x_N)$ is related to $x_a^i$ by
$x_a^i=\sum_{j=1}^N A_{i,j} x_j$.
Other elements ${\Phi}_j^{*}(z)$ $(j \neq 1)$ of the vertex
operators are determined by the intertwining property.
We conjecture that this bosonization of
the vertex operator ${\Phi}^{*}(z)$
gives the intertwiner 
among our bosonization and the typical representation.
Using the Gelfand-Zetlin basis \cite{PT},
we have checked this conjecture in some cases for $N=2,3,4$ \cite{Kojima3}.
We balance the "background charge" of the vertex operators
by using the screening ${\cal Q}_i$. 
Sometimes we have to multiply nontrivial product of
the screenings ${\cal Q}_i$ inside the vertex operator
in order to have non-zero correlation functions (trace of vertex operators).
At the end of this paper, we would like to give some comments 
on relating works \cite{Kojima2, Kojima0}.
In \cite{Kojima2} we study
how to get the Wakimoto realization from our bosonization by
$\xi$-$\eta$ system.
In \cite{Kojima0} we study
how to construct the elliptic deformed algebra 
$U_{q,p}(\widehat{sl}(M|N))$ from the quantum group
$U_q(\widehat{sl}(M|N))$.
Using deformation method developed in \cite{Kojima0} 
we obtain a bosonization of the elliptic deformed algebra 
$U_{q,p}(\widehat{sl}(N|1))$ for an arbitrary level $k$.

~\\
{\bf Acknowledgement}\\
This work is supported by the Grant-in-Aid for
Scientific Research {\bf C} (21540228)
from Japan Society for Promotion of Science.

\end{document}